 \documentclass[journal,twoside]{IEEEtran}

       \usepackage{soul}
 	\usepackage{graphicx}
	\graphicspath{{../pdf/}{../jpeg/}}
	\DeclareGraphicsExtensions{.pdf,.jpeg,.png..eps}

	\usepackage[cmex10]{amsmath}
	\usepackage{mathabx}
	\usepackage{algorithmic}
	\usepackage{array,multirow}
	\usepackage{mdwmath}
	\usepackage{eqparbox}
      \usepackage{arydshln}                
      \RequirePackage{doi}

   \usepackage{hyperref}
   \usepackage[normalem]{ulem}
         \usepackage{caption}
        \usepackage{subcaption}
    \newcolumntype{P}[1]{>{\centering\arraybackslash}p{#1}} 

\newlength{\Oldarrayrulewidth}

\usepackage{xcolor}  
\begin{document}

\title{Shape Anisotropy Enabled Field Free Switching of Perpendicular Nanomagnets}
\author{Akanksha Chouhan, Heston A. Mendonca, Abhishek Erram and Ashwin A. Tulapurkar
\thanks{All authors are with Department of Electrical Engineering, Indian Institute of Technology Bombay, Mumbai 400076, India  (e-mail: ecakanksha@ee.iitb.ac.in). }}

\maketitle

\begin{abstract}
Spin Orbit Torque-Magnetic Random Access Memory (SOT-MRAM)  is being developed as a successor to the Spin transfer torque MRAM (STT-MRAM) owing to its superior performance on the metrics of reliability and read-write speed. SOT switching of perpendicularly magnetized ferromagnet in the heavy metal/ferromagnet  bilayer of SOT-MRAM unit cell requires an additional external magnetic field to support the spin-orbit torque generated by heavy metal to cause deterministic switching. This complexity can be overcome if an internal field can be generated to break the switching symmetry. We experimentally demonstrate that by engineering the shape of ferromagnet, an internal magnetic field capable of breaking the switching symmetry can be generated, which allows for deterministic switching by spin-orbit torques. We fabricated nanomagnets of Cobalt with triangular shape on top of Platinum and showed external magnetic field free switching between the two stable states of magnetization by application of nano-second voltage pulses. The experimental findings are consistent with the micro-magnetic simulation results of the proposed geometry.
\end{abstract}

\IEEEoverridecommandlockouts
\begin{IEEEkeywords}
spin orbit torque, field free switching, symmetry engineering, shape anisotropy, SOT-MRAM
\end{IEEEkeywords}

\IEEEpeerreviewmaketitle


\section{Introduction}
In the growing data driven economy, MRAM is an emerging non-volatile (NVM) solution. Tremendous efforts are ongoing to make MRAM a viable option in the on-chip and off-chip memory segment. STT-MRAM has already found commercial presence due to its uniqueness of non-volatility and high endurance and its compatibility with CMOS fabrication process flow  \cite{worledge2024spin,ikegawa2020magnetoresistive}. However, the STT-MRAM suffers from reliability issue as a high write current passes through a very thin tunnel oxide in its magnetic tunnel junction (MTJ) unit cell  \cite{akerman2004demonstrated}. SOT-MRAM solves the STT-MRAM's reliability issue as SOT-MRAM's unit memory cell provides separate read and write current paths where the write current does not pass through the thin tunnel oxide. The SOT-MRAM unit cell is a three terminal device  \cite{liu2012spin}, where an MTJ is located over a bottom heavy metal (HM) layer such that the free layer of MTJ is in contact with the HM layer. The pMTJs in which free and pinned layers are perpendicularly magnetized are MTJs of choice for STT as well as SOT-MRAM as they provide more scalability, thermal stability and consume less power\cite{ikeda2010perpendicular}.

The HM/ferromagnet(FM) bilayer used in pMTJ SOT-MRAM is generally studied separately as HM/FM stack without integration of MTJ structure,  to understand SOT induced switching. In this stack, a charge current in $\hat{x}$ direction generates a $\hat{y}$ direction in-plane damping-like torque (DLT) which aligns the magnetization along $\hat{y}$ direction \cite{liu2012current}. When the charge current is switched off, the magnetization can relax to the $m_z=+1$ or $m_z=-1$ state with equal probability, thus making the switching non-deterministic. To obtain deterministic switching,  a small external symmetry breaking magnetic field is applied along the direction of charge current. To provide an external  magnetic field on-chip requires more complicated setup  and is therefore the major roadblock for the industry adoption of SOT-MRAM. This necessitates an external Field Free Switching (FFS) approach to deterministically switch magnetization. In the past multiple FFS approaches have been proposed \cite{wu2022field,ji2024recent}, such as - symmetry engineering to break the mirror symmetry by way of i) asymmetric device design which includes non-uniform film thickness \cite{yu2014switching,you2015switching}, non-uniform composition \cite{shu2022field,wu2020chiral}, asymmetric geometric shapes \cite{safeer2016spin}, canted SOT structures \cite{wang2019field,honjo2019first}, ii) crystal symmetry breaking \cite{liu2021symmetry,liu2022current}, iii) magnetic symmetry breaking \cite{bose2022tilted,fukami2016magnetization,ryu2022efficient}; 
compensated spin currents \cite{ma2018switching}, 
embedding in-plane magnet over pMTJ \cite{garello2019manufacturable} and many more.

\begin{figure}[ht!] 
\centering
\includegraphics[width= 3.0 in,keepaspectratio]{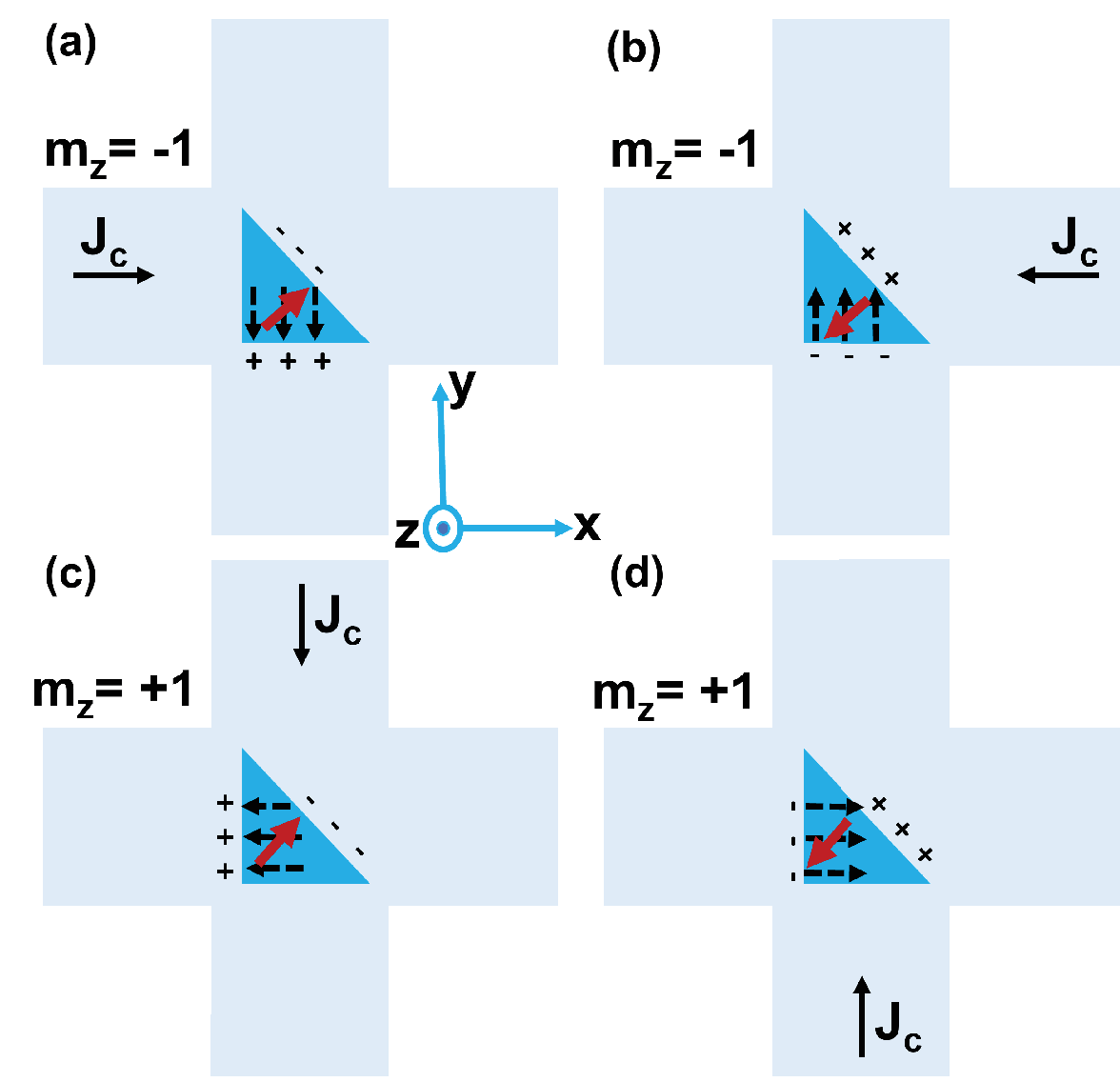}
\caption{Schematic of the proposed triangular shaped Co PMA nanomagnets over Pt Hall bar. The DLT arising from charge current in Pt aligns the magnetization in-plane, shown by the dotted black arrows. The magnetic charges induced on the periphery of triangle are denoted by plus and minus signs. The resultant demag field is shown by red arrow. The value of $m_z=\pm 1$ indicates the final magnetization state after current is switched off. (a), (b) the component of internal demag field along  $J_C$ is positive and magnetization relaxes to -$\hat{z}$ direction when the charge current is switched off. (c), (d) the component of internal demag field along  $J_C$ is negative and magnetization relaxes to +$\hat{z}$ direction when the charge current is switched off.}
\label{fig1}
\vspace{-0.2in}
\end{figure}

In this article, we experimentally demonstrate an FFS mechanism by utilizing asymmetric geometric shape of the perpendicular nanomagnet to break the mirror symmetry of switching. Switching through engineering shape anisotropy has been explored through simulations previously \cite{wang2019field}.  
 Fig.\ref{fig1} shows a Hall bar made of platinum (Pt) over which a perpendicularly magnetized cobalt (Co) nanomagnet with triangular shape is fabricated. Charge current flowing along $\hat{x}$ direction in Pt, produces a spin current polarized along -$\hat{y}$ direction which is absorbed by the Co. For a sufficiently large charge current density($J_c$), the resulting DLT 
 aligns the magnetization  along -$\hat{y}$ direction, and generates in-plane demag field as shown schematically in fig.\ref{fig1}a by dotted black and red arrow respectively. Demagnetization field in fig.\ref{fig1}a, has a component parallel to $J_c$. It causes a small tilt in magnetization along -$\hat{z}$ and when the current is switched off $m_z$=-1 state is stabilized \cite{liu2012current}. When the current is applied in -$\hat{x}$ (fig.\ref{fig1}b), DLT  aligns Co magnetization along +$\hat{y}$. Again the demag field has a component along $J_c$ and the magnetization relaxes to $m_z=-1$ state when the current is switched off. Thus, changing the current polarity does not change the final magnetization state. When charge current is passed along -$\hat{y}$, the magnetization is aligned along -$\hat{x}$ due to the DLT (fig.\ref{fig1}c), where as if the charge current is passed along +$\hat{y}$, the magnetization is aligned along +$\hat{x}$ (fig.\ref{fig1}d). In both these cases, the internal magnetic field has a component  opposite to the charge current and $m_z=+1$ state is reached after the current is switched off.

\begin{figure}[b] 
\vspace{-0.1in}
\centering
\includegraphics[width= 3.0 in,keepaspectratio]{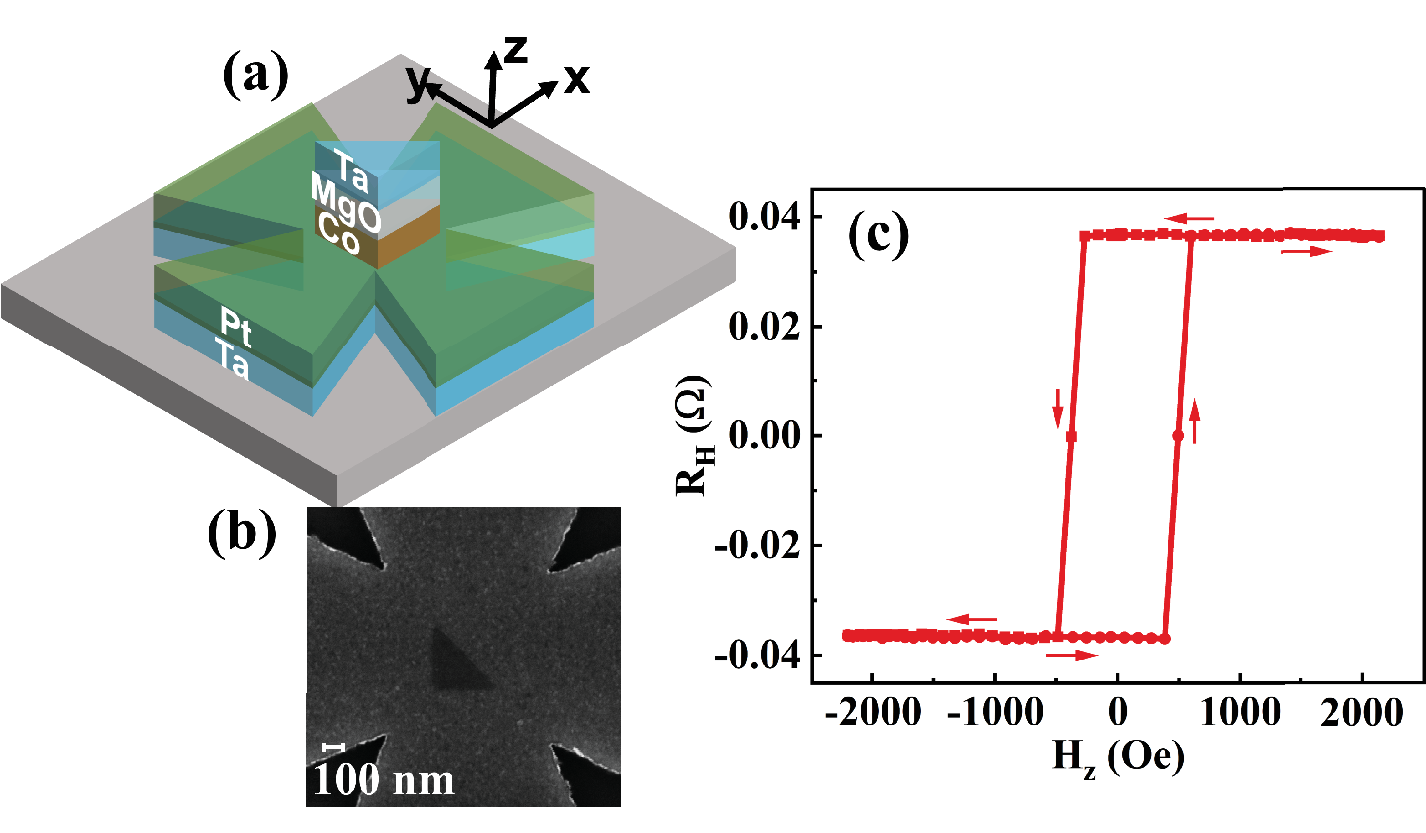}
\caption{(a) Schematic of the proposed device – triangular shaped Co PMA nanomagnet over Ta/Pt Hall bar (b) SEM image of the device (c) AHE hysteresis loop of the device}
\label{fig2}
\end{figure}

\begin{figure*} 
\centering
\vspace{-0.2in}
\includegraphics[width= 7.0 in,keepaspectratio]{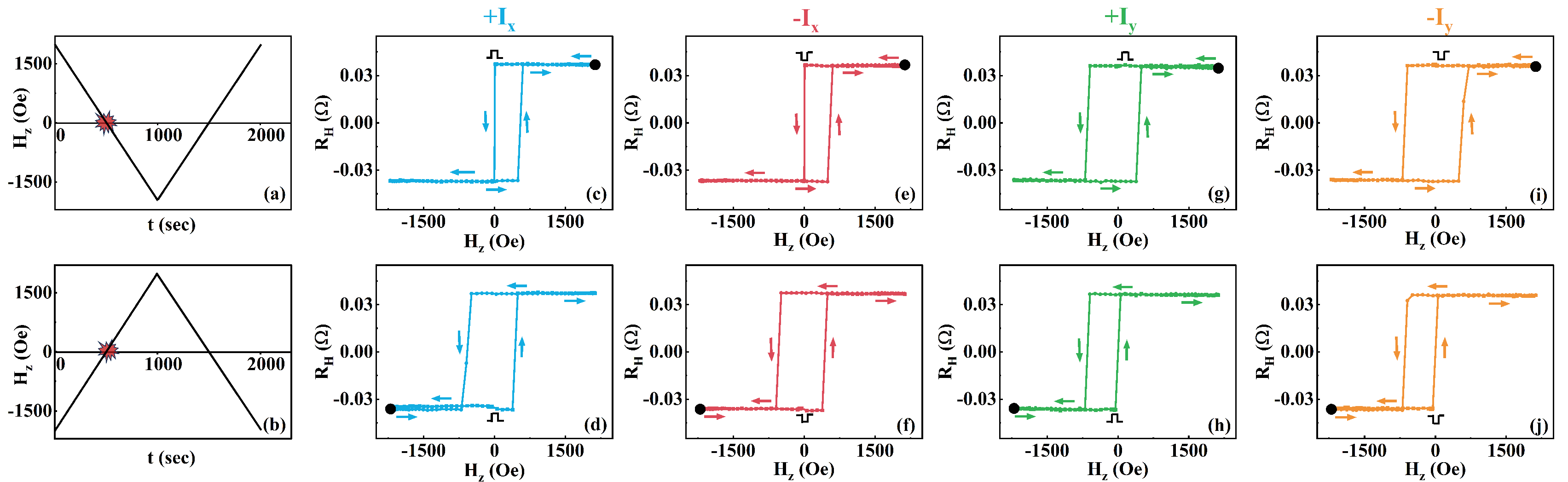}
\caption{Field Free Switching of magnetization of triangular shaped  devices, established through AHE measurements. (a), (b) The measurement sequences. AHE hysteresis loops are measured by passing charge current either along $\hat{x}$ or $\hat{y}$ direction, with magnetic field swept along $\hat{z}$ direction as indicated in (a) or (b). The star symbol denotes application of 10 $\mu s$ pulse with amplitude of 2 V or -2 V, when the external magnetic field is zero. (c) to (j): The black dot indicates starting point of the magnetic field sweep. The direction of current $\pm x/y$ is indicated on the top. The polarity of voltage pulse is also indicated in each panel. Fig (c)-(f) show that $m_z=-1$ state is reached when charge current is passed along $\hat{x}$ or -$\hat{x}$ direction. Fig (g)-(j) show that $m_z=+1$ state is reached when charge current is passed along $\hat{y}$ or -$\hat{y}$ direction.}
\label{fig3}
\vspace{-0.1in}
\end{figure*}

\begin{figure}[b!] 
\centering
\vspace{-0.4in}
\includegraphics[width= 3.5 in,keepaspectratio]{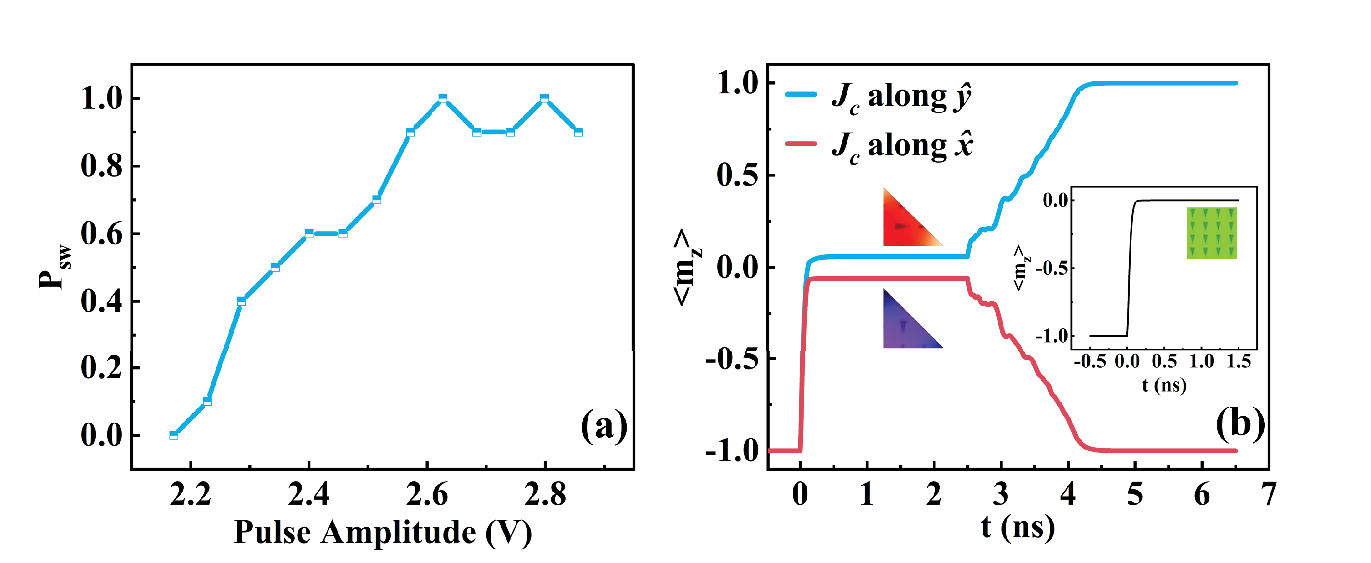}
\caption{(a) Switching Probability as a function of pulse amplitude for 10 ns pulses (b) Micromagnetic simulation of field free switching of triangular nanomaget due to SOT. Current along $\hat{y}$ leads to a state with $m_z>0$ when current is on. Magnetization switches to +$\hat{z}$ when the current is switched off.  Current along $\hat{x}$ results in a state with $m_z<0$ when current is on and magnetization switches to -$\hat{z}$ when the current is switched off. The inset shows micromagnetic simulation of square shaped FM. Application of current leads to a state with  $m_z=0$, indicating that magnetization can switch to $\pm \hat{z}$ with equal probability when the current is turned off.}
\label{fig4}
\end{figure}

\section{Results and Discussion}
For the FFS experiment, the Hall bar stack was deposited in two steps. At first, a Hall bar structure was patterned through e-beam lithography and then Ta(1)/Pt(5) were deposited, where numbers in the bracket represent thickness of the corresponding material in nm. Post lift-off, in the second step isosceles right-angle triangle structures with  250-400 nm sides were patterned at the centre of the Hall bar, through e-beam lithography. This was followed by the deposition of Co(1)/MgO(1.5)/Ta(1.5). Fig.\ref{fig2}a and \ref{fig2}b show the schematic and the SEM image of the triangular device respectively. The device has a $\Delta$R of 0.07 $\Omega$ and a switching field of 435 Oe as is shown in hall voltage ($R_H$) vs out-of plane field ($H_z$) anomalous hall effect (AHE) hysteresis loop in fig.\ref{fig2}c.

In order to demonstrate field free switching, we performed AHE hysteresis measurements as shown schematically in fig.\ref{fig3}a-b. For the measurement sequence shown in fig.\ref{fig3}a, we start with high magnetic field of 2000 Oe, which initializes the magnetization along $\hat{z}$. The charge current is passed either along the $\hat{x}$ or $\hat{y}$ and the anomalous Hall voltage is measured as a function of time while the magnetic field is ramped down and up. When the magnetic field crosses zero for the first time, a 10 $\mu$s voltage pulse with amplitude of either $+2V$ or $-2V$ is applied to the device. Whether the application of the voltage pulse resulted in magnetization switching or not can be figured out from the measurement of the entire hysteresis loop. The measurements as shown in fig.\ref{fig3}b were also performed where the magnetization is initialized along -$\hat{z}$. The results of these measurements are shown in fig.\ref{fig3}c-j. The dot in each sub-figure denotes the starting point ($m_z=1$ or $m_z=-1$). The polarity of the voltage pulse applied at zero magnetic field  is indicated in the figure. The direction of the pulse, whether the pulse was applied along the $\hat{x}$ or $\hat{y}$ is also indicated in the figure. In fig.\ref{fig3}c, when a positive voltage pulse was applied in the $\hat{x}$ direction to the device, the magnetization  switched from the state $m_z=+1$ to the state $m_z=-1$. Fig.\ref{fig3}d shows that, if we start with $m_z=-1$ and apply the same voltage pulse, the magnetization does not switch. Fig.\ref{fig3}e shows that starting from $m_z=+1$ state, application of negative voltage pulse along $\hat{x}$ also switches the magnetization to $m_z=-1$ state, where as fig.\ref{fig3}f shows that if we start from $m_z=-1$ state and apply the same negative pulse, magnetization does not switch. Thus, fig.\ref{fig3}c-f show that application of current either along +$\hat{x}$ or -$\hat{x}$ direction stabilizes $m_z=-1$ state without any assistance from external symmetry breaking magnetic field. In a similar way, the measurement results shown in fig.\ref{fig3}g-j show that application of current either along +$\hat{y}$ or -$\hat{y}$ direction stabilizes $m_z=+1$ state. 

Next the  experiments shown by measurement sequence in fig.\ref{fig3}a were carried out using nanosecond pulses. Voltage pulses of 10 ns duration with varying positive amplitude from 3.8 V to 5V were applied along $\hat{x}$ direction. At each amplitude point, measurements were carried out 10 times to find out the switching probability. The application of pulse at zero magnetic field is expected to switch the magnetization from $m_z=+1$ to $m_z=-1$ state.  For pulse amplitude upwards of 4.5V, a switching probability upwards of 90\% was registered as shown in fig.\ref{fig4}a.

 The experimental design was also tested using MuMax3 micromagnetic simulation framework \cite{Vansteenkiste2014}. 1 nm thick Co along with  a saturation magnetization ($M_s$) of 1.1$\times$$10^6$ A/m, damping coefficient ($\alpha$) of 0.1, uni-anisotropy ($K_u$) of 8.45$\times$$10^5$ J/$m^3$ and an exchange stiffness constant ($A_{ex}$) of 15$\times$$10^{-12}$ J/m  was used in the simulation. The simulations were carried out at zero temperature.
 The simulations were carried out for two different shapes of Co: an isosceles right angle triangle with 75 nm side and square with 75 nm side. 
Fig.\ref{fig4}b shows average value of the $\hat{z}$ component of magnetization  of triangular Co as a function of time when a 2.5 ns pulse with current density of 1.25$\times$$10^{13}$A/$m^2$  was passed through a Pt layer beneath. 
The spin Hall angle of Pt was taken as 0.07. When charge current density in Pt is along $\hat{y}$ (blue curve), the magnetization mostly points along $\hat{x}$ and the average value of $m_z$ is positive. When the current is turned off, the magnetization relaxes to +$\hat{z}$ direction. When the charge current density in Pt is along $\hat{x}$ direction, the magnetization mostly points along -$\hat{y}$ direction and the average value of $m_z$ is negative as shown by the red curve. In this case, when the current is turned off, the magnetization relaxes to -$\hat{z}$ direction. This is consistent with the  the experimental results. The inset shows the average value of the $\hat{z}$ component of magnetization of a squared shaped Co  as a function of time, when the charge current density in Pt is along $\hat{x}$ direction. The average value of $m_z$ in this case is 0 when the current is on, and thus magnetization can relax to $\pm \hat z$ direction with equal probability.

\section{Conclusion}
In conclusion, this work demonstrates deterministic switching of triangular shaped PMA nanomagnet without the assistance of an external symmetry breaking magnetic field. The shape anisotropy of the triangular structure resulted in an internal symmetry breaking field which was sufficient to cause deterministic SOT switching. This FFS solution is easier to implement
compared to other proposed methods involving thickness or composition gradient or requiring non-silicon substrates to grow material stacks for unconventional SOT. This proof-of-concept would pave way for further exploration for realizing SOT-MRAM using these structures.












\bibliographystyle{IEEEtran}

\bibliography{Ref}

\end{document}